\documentclass[lettersize,journal]{IEEEtran}
\usepackage{amsmath,amsfonts}
\usepackage{algorithmic}
\usepackage{array}
\usepackage[caption=false,font=normalsize,labelfont=sf,textfont=sf]{subfig}
\usepackage{textcomp}
\usepackage{stfloats}
\usepackage{url}
\usepackage{verbatim}
\usepackage{graphicx}
\def\BibTeX{{\rm B\kern-.05em{\sc i\kern-.025em b}\kern-.08em
    T\kern-.1667em\lower.7ex\hbox{E}\kern-.125emX}}
\usepackage{balance}
\begin{document}
\title{Maximization of Communication Network Throughput using Dynamic Traffic Allocation Scheme}
\author{Md.~Arquam, Suchi~Kumari,~\IEEEmembership{Senior~Member,~IEEE}         

\thanks{Md. Arquam is with the Department of Computer Science Engineering, IIIT Sonepat, Sonepat, Haryana, India. (Email: md.arquam@iiitsonepat.ac.in).

S. Kumari is with the Department of Computer Science and Engineering, Shiv Nadar School of Eminence, Delhi-NCR, India (Email: suchi.singh24@gmail.com).

}
}

\maketitle

\begin{abstract}
    Optimizing network throughput in real-world dynamic systems is critical, especially for diverse and delay-sensitive multimedia data types such as VoIP and video streaming. Traditional routing protocols, which rely on static metrics and single shortest-path algorithms, were unable in managing this complex information. To address these challenges, we propose a novel approach that enhances resource utilization while maintaining Quality of Service (QoS). Our dynamic traffic allocation model prioritizes different data types based on their delay sensitivity and allocates traffic by considering factors such as bandwidth, latency, and network failures. This approach is shown to significantly improve network throughput compared to static load balancing, especially for multimedia applications. Simulation results confirm the effectiveness of this dynamic method in maximizing network throughput and maintaining QoS across various data types.
\end{abstract}


\section{Introduction} \label{sec1}
Real-world networks, including communication, rail, electrical, and transportation systems, are complex and constantly evolving. With the advancement in communication technology, it is required to satisfy the requirement of network traffic to maintain Quality of Service (QoS) with improved network resource utilization and enhanced throughput. In traditional routing protocols, a single shortest path is used to send data through a source node to a destination node based on statically configured metrics \cite{kumari2019fair}. Several algorithms exist for finding the shortest path between desired nodes in a given network \cite{yu1998robust,arquam2014delay}. However, these algorithms are designed for deterministic networks and can not be applied to real work networks. In traditional routing protocols, network load were statistically distributed among the nodes within the network and resources were underutilized due to the lack of selection of efficient load balancer technique for distributing loads among various applications. However, the algorithms to solve such a problem have been shown to have, in general, high computational complexity. Therefore, in this paper, a novel approach is provided to enhance network throughput depending on the demand of network resources by various applications. The traffic is dynamically distributed considering various metrics such as available bandwidth, latency sensitivity, QoS, priority of the traffic and so on.

Researchers evaluated the network throughput considering wired as well as wireless network framework. Curado \textit{et al.} \cite{curado2004survey} surveyed multiple approaches for the selection of Constraint shortest paths such as bandwidth-restricted path \cite{sobrinho2002algebra}, delay constrained path \cite{vyas2017dhac}, weighted sum of link matrix \cite{juttner2001lagrange}, and so on. Some researchers evaluated the QoS of wireless networks such as wireless multimedia sensor networks (WMSNs)\cite{li2019survey}, underwater wireless sensor networks (UWSNs) \cite{luo2021survey}, mobile ad hoc networks (MANETs) \cite{zhang2019kind}, etc.

In a large network, a node with a larger degree gets easily congested as a large number of traffic is received from all the adjacent nodes. Therefore, an efficient routing strategy avoids the usage of a larger degree in the route selection from the source node to the destination node in the network. Wang \textit{et al.} \cite{wang2011dynamic} divided the network traffic into multiple sub-traffics and data transmission was done parallel to minimize delay and communication cost. In \cite{bai2019multi}, a routing algorithm is provided by assigning a priority score based on its source node.  Liu \textit{et al.} \cite{liu2020cost} minimized the communication cost by routing the traffic through multiple paths considering maximum delay constraint and throughput. The approximation algorithm and an efﬁcient heuristic approach leverage edge-cloud computing platforms to support computationally intensive IoT applications.

From the literatire survey, it can be analyzed that the network throughput is optimized by assigning more weightage to constrained path selection \cite{salama1997distributed}, multi-path selection \cite{liu2020cost}, secure connection establishment \cite{kumari2021intelligent}, etc., to find a reliable communication path from the source node to the destination node to ensure the QoS of the system. More or less, the path/s is/are predefined based on network metrics before transmitting user data, although throughput is influenced by dynamic parameters such as runtime delay, network failures, dynamic bandwidth requirements, packet loss, jitter, etc. Apart from that, the type of data transmitted across the network can include multimedia data such as Voice over IP (VoIP), video streaming, web browsing, and file downloads. These data types share a single channel, and their bandwidth requirements vary accordingly. Delay sensitivity also differs; for instance, video streaming and VoIP are highly sensitive to delay and therefore need higher priority. Considering these factors, this paper proposes a model for dynamic traffic allocation that accounts for various dynamic parameters and assigns distinct priorities to different types of applications, addressing both traffic and network complexity. The detailed objective is as follows.
\begin{itemize}
    \item Propose a novel method to enhance network throughput by considering the resource demands of various applications.
    \item Assign priority to different types of multimedia data based on their delay sensitivity scores.
    \item Analyze the QoS scores for various data types using a dynamic traffic allocation scheme.
    \item Compare the impact of a dynamic load balancer versus a static load balancer on network throughput for multimedia data types.
\end{itemize}

The rest of the paper is organized as follows: Section \ref{sec1} provides the introduction, literature survey and recent trends in the traffic allocation in the network. Section \ref{sec3} presents the strategies for the network throughput maximization for multimedia data. Section \ref{sec4} demonstrates the results and analysis of network throughput estimation considering dynamic traffic allocation (DTA) scheme. Section \ref{sec5} discusses the conclusions and future scope of the work.
\section{Proposed Methodology} \label{sec3}
In this section, a novel method is proposed to enhance the network throughput considering various parameters such as types and priority of the applications, usage of the network resources, current network traffic, disturbances or jitter in the network and so on. The assignment of priority increases the network throughput depending upon demand of resources by applications. Because each and every application do never use 100\% resource and sometimes resource remains unused. It needs to be optimized to maximize communication network throughput in order to achieve efficient and trustworthy network performance. Hence, Dynamic Traffic Allocation scheme is provided for the network resource distribution based on current demand and circumstances. To formulate the model, various parameters are used so the notations and its description is provided in Table \ref{tab:my_label}.

\begin{table}[!htb]
    \centering
    \caption{Notations and their meaning}
    \begin{tabular}{|p{1.25 cm}|p{6.5 cm}|}
    \hline
      \textbf{Notations} & \textbf{Meaning} \\ \hline
      $Tr(t)$   &  network throughput at time $t$\\ \hline
      $A_i(t)$ & Allocated resources for traffic type $i$ at $t$.\\ \hline
      $D_i(t)$ & Resource demand for $i$ at time $t$.\\ \hline
      $R(t)$ & Resources available at time $t$.\\ \hline
      $P_i(t)$ & Traffic type $i$'s priority at time $t$.\\ \hline
      $B_i(t)$ & Requirements for bandwidth for $i$ at time $t$.\\ \hline
      $Lt_i(t)$ & Sensitivity to latency for $i$ at time $t$.\\ \hline
      $QoS_i(t)$ & Requirement for QoS for $i$ at time $t$.\\ \hline
      $Ld_i(t)$ & Load on resource $i$ at time $t$.\\ \hline
      $Jtr_i(t)$ & Maximum acceptable jitter for  $i$ at time $t$.\\ \hline
      $Pkt_i(t)$ & Packet loss ratio for $i$ at time $t$.\\ \hline
    \end{tabular}
    
    \label{tab:my_label}
\end{table}

The network throughput $Tr(t)$ can be evaluated by summing the overall resource allocation done for each traffic type $i$ at time $t$. The formulation is provided in Eq. \eqref{throughput}.
\begin{equation}
\begin{aligned}
Tr(t)=\sum_i^N A_i(t) \\
A_i(t)=f_i(t)T(t).  \\  \label{throughput}
\end{aligned}
\end{equation}
Where $f_i(t)$ represents the fraction of the  traffic contributed by application $i$ at time $t$. The goal is to maximize throughput while keeping resource availability and demand in mind. This is expressed as an optimization problem considering various constraints such as latency, jitter and packet loss (in Eq. \eqref{maxtr}).

\begin{equation}
\begin{aligned}
\max Tr \quad &  \sum_{i=1}^{N} P_i(t).\frac{A_i(t)}{D_i(t)} \\
\textrm{s.t.} \quad & 0 \leq A(t) \leq R(t) \\
&A_i(t) \geq B_{Demanded,i}    \\
& Lt_i(t)\leq  Lt_{max,i}    \\
&Jtr_i(t)\leq  Jtr_{max,i}    \\
&Pkt_i(t)\leq  Pkt_{max,i}    \\ \label{maxtr}
\end{aligned}
\end{equation}
Now we need to incorporate the concept of dynamic resource allocation, which takes into account various factors such as \textit{bandwidth requirements}, \textit{latency sensitivity}, \textit{Quality of Service (QoS)}, and the \textit{priority} of different types of traffic.

The allocated resource, $(A(t))$, is dynamically adjusted in response to changing traffic patterns. Resources are assigned to each traffic type based on its priority relative to the overall priority distribution. High-priority traffic is allocated a larger share of the available resources, ensuring that the allocation does not exceed the bandwidth requirements of each traffic category. Resource allocation is also influenced by the latency sensitivity and demand of each traffic type, with more latency-sensitive traffic and demand receiving a larger portion of the resources. Resource allocation is also proportional to the relative Quality of Service (QoS) needs of each traffic type in relation to the overall QoS requirements.

\begin{equation}
\begin{aligned}
A_i(t) \propto \quad &   \frac{P_i(t)}{\sum_{j} P_j(t)} \\
\approx  \quad & min(B_i(t),R(t)) \\
\propto \quad & \frac{Lt_i(t)}{\sum_{j} Lt_j(t)}  \\
\propto \quad &  \frac{D_i(t)}{\sum_{j} D_j(t)}    \\
\propto \quad & \frac{QoS_i(t)}{\sum_{j} QoS_j(t)}    \\
\end{aligned}
\end{equation}
The QoS is measured using Eq. \eqref{QoS}.
\begin{equation}
    QoS_i(t)=\frac{B_i(t)}{B_{(Demanded,i)}}.\frac{Lt_{(max, i)}}{Lt_i(t)}.\frac{Jtr_{(max,i)}}{Jtr_i(t)}.(1-Pkt_i(t)) \label{QoS}
\end{equation}

\subsection*{Load balancing}
Load balancing ensures the equitable distribution of traffic across network resources for optimal utilization. In a dynamic resource allocation system, both static and dynamic load balancing algorithms can be utilized. The following equations provide the mathematical representation for evaluating static and dynamic traffic.
\begin{itemize}
    \item \textbf{Static Load Balancing:} The resource is equally distributed among all the traffic type $i \in N$ and represented in Eq. \eqref{static}. 
    \begin{equation}
        A_i(t)\propto (1/N) \label{static}
    \end{equation}
    \item \textbf{Dynamic Load Balancing:} The allocation is adjusted dynamically based on the current load on each traffic type $i$. It redistributes resources to less heavily laden resources and considers the demand for each traffic type.
    $$A_i(t) \propto \frac{1-Ld_i(t)}{\sum_{j}(1-Ld_i(t))}.\frac
    {(D_i(t)}{\sum_{j}D_j(t))}$$
\end{itemize}
The dynamic load balancing algorithm considers both resource load and resource demand. The load $Ld_i(t)$ can encompass metrics such as CPU usage, bandwidth utilization, or other relevant indicators. The following equation (Eq. \eqref{dynamiclb}) incorporate these factors based on network and traffic trends.

\begin{equation}
\begin{aligned}
A_i(t) = \quad &  R(t).\frac{P_i(t)}{\sum_{j} P_j(t)}.\frac{(min(Bi(t),R(t))}{B_i(t))}. \frac{Lt_i(t)}{\sum_{j}Lt_j(t)}. \\
&\frac{D_i(t)}{\sum_{j} D_j(t)}.\frac{QoS_i(t)}{\sum_{j} QoS_j(t)}.\frac{1-Ld_i(t)}{\sum_{j}(1-Ld_i(t))}  \\
& + R(t).\frac{1}{N} \label{dynamiclb}
\end{aligned}
\end{equation}

The $A_i(t)$ score is provided to Eq. \eqref{throughput} to evaluate the network throughput for the real communication network. 

\section{Results and Analysis} \label{sec4}
In this section, a communication network of $50$ nodes is created with various network parameters. The detailed score is provided in Table \ref{tab:parameter}.
\begin{table}[!htb]
    \centering
    \caption{Network parameters with the assigned values}
    \begin{tabular}{|p{3.5 cm}|p{3 cm}|}
        \hline
         Parameter &  Range of values \\ \hline
         $N, D, R$ & $50, 2, 50$ \\ \hline
         $P, P_{max}$ & $0.5, 0.01$ \\ \hline
         $Lt, Lt\_max, Lt\_range$  & $30, 50, (10, 100, 100)$ \\ \hline
         $Jtr, Jtr\_max, Jtr\_range$ & $2, 5, (0.1, 10, 100)$ \\ \hline
         $B\_required$ & $10$ \\ \hline
         Total Traffic types & 4 \\ \hline
        $packet\_loss\_range$  & (0, 0.05, 100)\\ \hline
          $bandwidth\_range$ & (5, 15, 100)\\ \hline
    \end{tabular}
    \label{tab:parameter}
\end{table}

\begin{figure}[!htb]
\begin{center}
\includegraphics[width=.7\linewidth,height=1.75in]{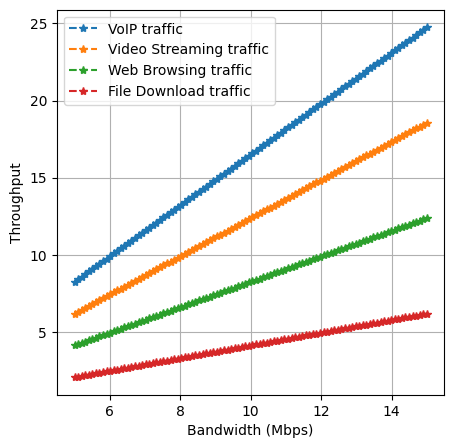}
\caption{The impact of Dynamics Traffic Allocation Scheme on network throughput for various types of network traffic.}
\label{Through_traffic}
\end{center}
\end{figure}

The impact of dynamic bandwidth allocation on throughput is illustrated in Figure \ref{Through_traffic}. The figure demonstrates how different types of traffic, each with varying priorities, affect efficiency as bandwidth increases. For example, VOIP traffic receives more bandwidth than other types of traffic, while File Download traffic uses less. The allocation is based on the priority of each application, with higher-priority applications achieving better throughput. It shows how the dynamic bandwidth sharing model prioritizes and optimizes resource distribution for superior applications.

\begin{figure}[!htb]
\begin{center}
$\begin{array}{cc}
\includegraphics[width=.45\linewidth,height=1.5in]{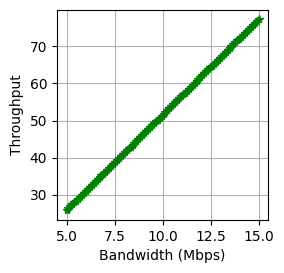}  &
\includegraphics[width=.45\linewidth,height=1.5in]{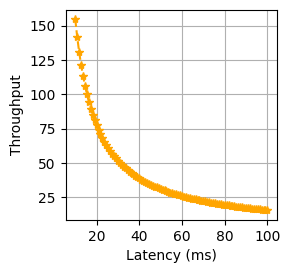} \\
\mbox{\small(a)} & \mbox{\small(b)} \\
\includegraphics[width=.45\linewidth,height=1.5in]{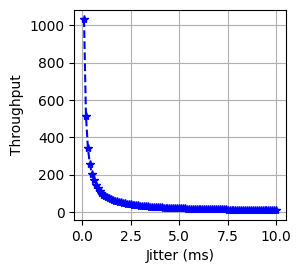}  &
\includegraphics[width=.45\linewidth,height=1.5in]{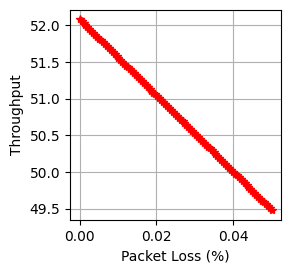} \\
\mbox{\small(c)} & \mbox{\small(d)} \\
\end{array}$
\caption{Plot showing impact of QoS parameters; (a) bandwidth requirement, (b) varying latency, (c) varying jitter and (d) packet loss rate, on throughput in Dynamics Traffic Allocation Scheme}
\label{Through_Qos}
\end{center}
\end{figure}

The impact of QoS parameters on throughput in a DTA scheme is depicted in Figure \ref{Through_Qos}. The plot shows that increasing bandwidth  enhances throughput (in Fig. \ref{Through_Qos}(a)), while increases in latency and jitter negatively affect throughput (in Fig. \ref{Through_Qos}(b) and in Fig. \ref{Through_Qos}(c), respectively). The relationship between latency, jitter, and throughput approximates a power law in the DTA scheme. The packet loss has an inversely proportional relationship with throughput, meaning that higher packet loss leads to a decrease in throughput (in Fig. \ref{Through_Qos}(d)).

\begin{figure}[!htb]
\begin{center}
\includegraphics[width=.8\linewidth,height=2.5in]{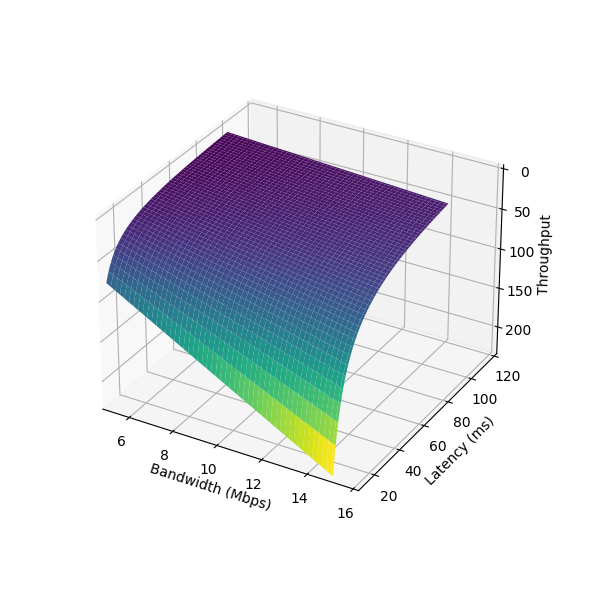}  \\
\caption{3-D Plot showing impact of bandwidth and Latency parameters on throughput in Dynamics Traffic Allocation Scheme}
\label{Through_3D}
\end{center}
\end{figure}

From the graph displayed in Fig. \ref{Through_3D}, we can infer how throughput changes with varying bandwidth and latency. It is observed that bandwidth has a more favorable impact on throughput compared to latency, which typically has a negative effect. This visualization helps clarify the correlation between these parameters.

\begin{figure}[!htb]
\begin{center}
$\begin{array}{cc}
\includegraphics[width=.45\linewidth,height=1.5in]{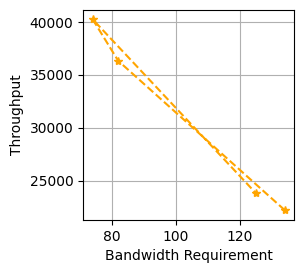}  &
\includegraphics[width=.45\linewidth,height=1.5in]{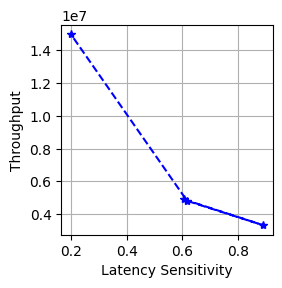} \\
\mbox{\small(a)} & \mbox{\small(b)} \\
\includegraphics[width=.45\linewidth,height=1.5in]{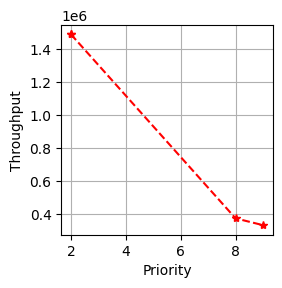}  &
\includegraphics[width=.45\linewidth,height=1.5in]{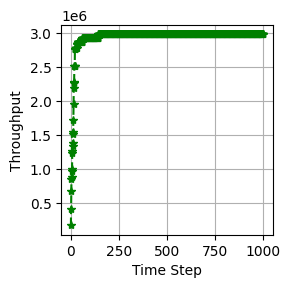} \\
\mbox{\small(c)} & \mbox{\small(d)} \\
\end{array}$
\caption{Plot showing Change in throughput according to (a) bandwidth requirement, (b) varying latency, (c) varying priority and (d) varying time in Dynamics Traffic Allocation Scheme}
\label{Through_var}
\end{center}
\end{figure}

The analysis of throughput under varying conditions, including bandwidth (Fig. \ref{Through_var} (a)), latency (Fig. \ref{Through_var} (b)), priority (Fig. \ref{Through_var} (c)), and time (Fig. \ref{Through_var} (d)), is illustrated in Figure \ref{Through_var}. As the bandwidth requirements increase within the DTA scheme, throughput efficiency declines, suggesting that higher bandwidth demands adversely affect throughput. Similarly, as latency sensitivity increases, throughput decreases, with higher sensitivity leading to reduced efficiency (Fig. \ref{Through_var} (b)). In Figure \ref{Through_var} (c), the graph appears linear, indicating that as priority increases, throughput decreases correspondingly. This suggests that higher task priorities impose stricter resource requirements, thereby reducing overall throughput efficiency. Over time, throughput initially experiences a rapid rise before stabilizing around 1.3 million (Fig. \ref{Through_var} (d)), demonstrating the DTA scheme's ability to quickly adapt and reach a stable, optimal state. This sharp initial increase highlights the network's adaptability and the system's capacity to achieve and maintain efficiency and stability as time progresses.


\section{Conclusions and Future Scope} \label{sec5}
In this paper, a novel method is proposed for enhancing network throughput by dynamically allocating resources based on various parameters such as application types, priorities, network resource usage, current traffic conditions, and jitter. The proposed DTA scheme optimizes resource distribution to maximize network throughput, considering factors like bandwidth requirements, latency sensitivity, QoS, and traffic demand. From the results, we can infer that the throughput is inversly proportional to the bandwidth demand, latency sensitivity and task prioritization. Due to the dynamic load balancing, DTA scheme quickly reach to a stable optimal state.

The future potential of dynamic traffic allocation is to extend the capabilities of the system that manages it by adding AIML, edge computing, SDN, and network slicing. Communication networks are always evolving, so researchers must seek maximum throughput and service quality from complex communication networks.

\bibliographystyle{IEEEtran}
\bibliography{manuscript}

\end{document}